\documentclass{ws-p8-50x6-00}

\begin{document}

\title{Magnetised Neutron Stars : An Overview}

\author{Ashok Goyal}

\address{Department of Physics \& Astrophysics,\\
University of Delhi, \\
Delhi - 110 007, India. \\
agoyal@ducos.ernet.in}


\maketitle

\abstracts{In the presence of strong magnetic field reported to have
been observed on the surface of some neutron stars and on what are
called Magnetars, a host of physical phenomenon from the birth of a
neutron star to free streaming neutrino cooling phase will be
modified. In this review I will discuss the effect of magnetic field
on the equation of state of high density nuclear matter by including
the anomalous magnetic moment of the nucleons into consideration. I
would then go over to discuss the neutrino interaction processes in
strong as well as in weak magnetic fields. The neutrino processes are
important in studying the propagation of neutrinos and  in studying the
energy loss, Their study is a prerequisite for the understanding of
actual dynamics of supernova explosion and on the stabilization of
radial pulsation modes through the effect on bulk viscosity. The
anisotropy introduced in the neutrino emission and through the
modification of the shape of the neutrino sphere may explain the
observed pulsar kicks.}

\section{Introduction}

Large magnetic fields $\sim 10^{14}$ Gauss have been reported to exist on
the surface of pulsars. Recent observations {\cite{Kouv}} of
$\gamma$-ray repeaters 
and spinning X-ray pulsars (Magnetars) hints to the existence of fields in 
excess of $10^{14}$ Gauss. It then follows from the Scalar Virial theorem
that the fields in the core could even reach a value as large as $10^{18}$
Gauss. There is however, an upper limit on the magnetic field discussed by
Chandershekhar beyond which the magnetic energy exceeds the
gravitational energy and the star is no longer stable.In the presence of such
 magnetic fields, neutron star properties in all its phases from its evolution
 from proto-neutron star to cold neutrino emitting phase would be modified.
 This arises because in the presence of magnetic field motion of the charged 
particles is quantised in a plane perpendicular to the magnetic field and the 
charged particles occupy discrete Landau levels. This has the effect of not
 only modifying the energy eigenvalues but also the particle wave functions.The
 quantum state of a particle in magnetic field is specified by momentum
 components $p_z,p_y$, spin s and Landau quantum number $\nu$. The anomalous
 magnetic moments of protons and neutrons further modify the energy
 eigenvalues. The time scales involved during all phases of neutron stars from
 birth to neutrino burst through thermal neutrino emission from trapped
 neutrino sphere to freely streaming neutrino cooling phase are large compared
 to the interaction time scales of strong, electromagnetic and weak
 interactions and the matter is in $\beta$ equilibrium. The magnetic field 
would modify the equilibrium and all neutrino interaction processes including
 scattering, absorption and production.

The stategy then is to first solve the Dirac equation for all particles in
 magnetic field including their anamolous magnetic moments, obtain the energy
 eigenvalues, construct the Grand Partition Function taking into account
 strong interactions in some model dependent way and obtain the equation of
 state (EOS). The next step is to calculate the scattering cross-sections for
 all neutrino processes by using the exact wave functions and by modifying the
 phase space integrals for arbitrary values of degenracy, density, temperature
 and magnetic field. The various phenomenon that I will address are :
\begin{enumerate}
\item{} Composition of matter in neutron stars, proton fraction,
effective nucleon  mass etc. 
\item{} Cooling of neutron stars in the free streaming regime. 
\item{} Neutrino transport in neutron and collapsing stars which is an
essential prerequisite for an understanding of supernova explosion,
structure of proto-neutron star and observed pulsar kicks. 
\item{} Damping of radial oscillations and secular instability through
the calculation of bulk viscosity.
\end{enumerate}

\section{Nuclear Matter Composition}

For determining the composition of dense, hot, magnetised matter, 
we employ a relativistic mean field theoretical approach in 
which the baryons (protons and neutrons) interact via the exchange of 
$\sigma-\omega-\rho$ mesons 
in a constant uniform magnetic field. Following reference {\cite{Agoyal}}
 in a uniform magnetic field B along z axis corresponding to the choice of the
gauge field $A^{\mu}=(0,0,xB,0)$,the relativistic mean field Lagrangian can be
written as
\begin{eqnarray}
{\cal L}
&=&
 \sum \bar{\psi_{B}}\bigg[i\gamma_{\mu}D^{\mu}-m_{B}+g_{\sigma B}\sigma
    -g_{\omega B}\gamma_{\mu}\omega^{\mu}
    -g_{\rho B}\,\tau _{3 B}\,\gamma_{\mu}\,\rho^{\mu}
	+ \kappa_{B} \sigma_{\mu \nu} F^{\mu \nu}\bigg]\psi_{B}  \nonumber \\
&&   +\frac{1}{2}\bigg[(\partial^{\mu}\sigma)^2-(m_{\sigma}^2\sigma^2)\bigg] 
  -\frac{1}{4}\omega_{\mu \nu}\omega^{\mu \nu}
	+\frac{1}{2}m_{\omega}^{2}\omega_{\mu}\omega^{\mu}
	-\frac{1}{4}\rho_{\mu \nu}\rho^{\mu \nu}     \nonumber \\
&&	+\frac{1}{2}m_{\rho}^{2}\rho_{\mu}\rho^{\mu}-U(\sigma)	
	-\frac{1}{4}F^{\mu \nu}F_{\mu \nu}
	+\sum_{l}\bar{\psi_{l}}(i\gamma_{\mu}D^{\mu}-m_{l})\psi_{l}
\end{eqnarray}
\noindent  in the usual notation with $D^{\mu}=\partial^{\mu}+ieA^{\mu}$ and 
$\kappa_{B}$ as the
anomalous magnetic moments given by
\begin{equation}
\kappa_{p}=\frac{e}{2m_{p}}\bigg[\frac{g_{p}}{2}-1\bigg]
 \ \ ;  \kappa_{n}=\frac{e}{2m_{n}}\frac{g_{n}}{2}
\end{equation}
\noindent where $g_{p}=5.58$   and $g_{n}=-3.82$  are the Lande's g-factor 
for protons and neutrons respectively. 
 Replacing the meson
fields in the relativistic mean field approximation by their density dependent
average values $<\sigma_{0}>$,$<\omega_{0}>$ and $<\rho_{0}>$, the 
equations of motion satisfied by the nucleons in the magnetic field become 
\begin{eqnarray}
 &&{} [-i\alpha_{x}\frac{\partial}
	{\partial_{x}}+\alpha_{y}(p_{y}-eBx) 
	+\alpha_{z}p_{z}+\beta(m_{i}-g_{\sigma}<\sigma_{0}>)-i\kappa_{i}
	\alpha_{x}\alpha_{y}B]f_{i,s} \nonumber  \\
&&{}= (E^{i}-U_{0}^i)f_{i,s}(x) 
\end{eqnarray}
 where
\begin{equation}
U_{0}^{p,n}=g_{\omega N}<\omega_{0}> \pm \frac{g_{\rho N}}{2}<\rho_{0}>
\end{equation}
The equations  are first solved for the case when momentum along the magnetic
field direction is zero and then boosting along that direction till the
momentum is $p_{z}$. For the neutrons and protons we thus get
\begin{eqnarray}
{\cal E}_{s}^{n} &=&
\sqrt{m_{n}^{*2}+\overrightarrow{p^{2}}+\kappa_{n}^2 B^2 + 2\kappa_{n}
B s\sqrt{p_{x}^{2}+p_{y}^{2}+m_{n}^{*2}}} \nonumber \\ &{}& =
E_{s}^{n}-U_{0}^{n}
\end{eqnarray}
and 
\begin{eqnarray}
 {\cal E}_{\nu,s}^{p} &=& \sqrt{m_{p}^{*2}+p_{z}^2+eB(2\nu+1+s)
+\kappa_{p}^2 B^2+2\kappa_{p} B
s\sqrt{m_{p}^{*2}+eB(2\nu+1+s)}} \nonumber \\ 
&=& E_{\nu,s}^{p}-U_{0}^{p}
\end{eqnarray}
\noindent where $2\nu=(2n+1+s)$, where $\nu$ and n being integer known as 
Landau 
quantum number and principal quantum number respectively. $s=\pm 1$ 
indicates whether the spin is along or opposite to the direction 
of the magnetic field and $m_{B}^*=m_{B}-g_{\sigma}\sigma$ is the effective 
baryon mass.
The energy spectrum for electrons is given by 
 \begin{equation}
E_{\nu,s}^{e}=\sqrt{m_{e}^2+p_{z}^2+(2\nu+1+s)eB}
\end{equation}
\noindent The mean field values $<\sigma_{0}>$, $<\omega_{0}>$ and 
$<\rho_{0}>$ are 
determined by minimizing  the energy at fixed baryon density
$n_{B}=n_{p}+n_{n}$ or by  
maximizing the pressure at fixed baryon chemical potential $\mu_{B}$. 
We thus get
\begin{equation}
<\sigma_{0}>=\frac{g_{\sigma}}{m_{\sigma}^{2}}(n_{p}^{s}+n_{n}^{s})
~,
<\omega_{0}>={\frac{g_{\omega}}{m_{\omega}^2}}(n_{n}+n_{p})
~,
<\rho_{0}>=\frac{1}{2}\frac{g_{\rho}}{m_{\rho}^{2}}(n_{p}-n_{n})
\end{equation}
\noindent  where $n_{i}$ and $n_{i}^{s}$ (i=e,p) are the number and
scalar number  
densities for proton and neutron.
\noindent In the presence of the magnetic field phase space volume is 
replaced as 
\begin{equation}
\int \frac{d^3p}{(2\pi)^3}\longrightarrow \frac{eB}{(2\pi)^2} \sum_{\nu=0}^{\nu_{max}} (2-\delta_{\nu,0}) \int \,dp_{z}
\end{equation}
\noindent The expressions for number densities, scalar densities for neutron 
and proton are given by
 \begin{eqnarray}
n_{n}&=&\frac{1}{(2\pi)^{3}}\sum_{s}\int 
	\frac{d^3p}{1+e^{\beta({\cal E}_{s}^{n}-\mu_{n}^{\star})}}  \\
n_{n}^{s} &=&\frac{m_{n}^{\star}}{(2\pi)^{3}}\sum_{s}\int 
	\frac{d^3p}{{\cal E}_{s}^{n}
	(1+e^{\beta({\cal E}_{s}^{n}-\mu_{n}^{\star})})}  \\
n_{p}&=&\frac{eB}{2\pi^{2}}\sum_{s}\sum_{\nu}\int_{0}^{\infty} 
	\frac{dp_{z}}{1+e^{\beta({\cal E}_{\nu,s}^{p}-\mu_{p}^{\star})}} \\
n_{p}^{s}&=&\frac{eB}{2\pi^{2}}m_{p}^{\star}\sum_{s}\sum_{\nu}\int 
	\frac{dp_{z}}{{\cal E}_{\nu,s}^{p}(1+e^{\beta({\cal
	E}_{\nu,s}^{p}-\mu_{p}^{\star})})} 
\end{eqnarray}
\noindent and the net electron and neutrino number densities are given by
\begin{equation}
\bar n_{e}=\frac{eB}{2\pi^{2}}\sum(2-\delta_{\nu,0})\int 
	dp_{z}\Bigg[\frac{1}{(1+e^(\beta({\cal E}_{\nu}^{e}-\mu_{e})))}
	-\mu_{e}\leftrightarrow(-\mu_{e})\Bigg]
\end{equation}
\begin{equation}
\bar n_{\nu_{e}}=\frac{1}{(2\pi)^{3}}\int 
	d^3p\Bigg[\frac{1}{(1+e^{\beta({\cal E}^{\nu_{e}}
	-\mu_{\nu_{e}})})}-\mu_{\nu_{e}}
	\leftrightarrow(-\mu_{\nu_{e}})\Bigg]
\end{equation}
 \noindent The thermodynamic potential for the neutron, proton,
electron and neutrino are given by  
\begin{equation}
\Omega_{n}=-\frac{1}{\beta(2\pi)^{2}}\sum_{s}\int 
	d^3p\ln[1+e^{-\beta({\cal E}_{s}^{n}
	-\mu_{n}^{\star})}]   
\end{equation} 
\begin{equation}
\Omega_{p}=-\frac{1}{\beta(2\pi)^{2}}\sum _{s}\sum_{\nu}\int 
	dp_{z}\ln[1+e^{-\beta({\cal E}_{\nu,s}^{p}
	-\mu_{p}^{\star})}] 
\end{equation} 
\begin{equation}
\Omega_{e}=-\frac{1}{\beta(2\pi)^{2}}\sum_{s} \sum_{\nu} \int 
	dp_{z}\ln[1+e^{-\beta ({\cal E}_{\nu,s}^{e}
	-\mu_{e})}] 
\end{equation} 
\begin{equation}
\Omega_{\nu_{e}}=-\frac{1}{\beta(2\pi)^{2}}\int 
	d^3p\ln[1+e^{-\beta({\cal E}^{\nu_{e}} 
	-\mu_{\nu_{e}})}] 
\end{equation} 
and the thermodynamic potential for the system is given by
\begin{equation}
\Omega=-\frac{1}{2}m_{\omega}^{2}\omega_{0}^{2}
	-\frac{1}{2}m_{\rho}^{2}\rho_{0}^{2}
	+\frac{1}{2}m_{\sigma}^{2}\sigma^{2}+U(\sigma)
	+\frac{B^{2}}{8\pi}+\sum_{i}\Omega_{i}
\end{equation}
\noindent where $i=n,p,e, \nu_{e}$.
 The various chemical potential are 
determined by the conditions of charge neutrality and  chemical equilibrium. 
In later stages of core collapse and during the early 
stages of protoneutron star, neutrinos are trapped and 
the chemical potentials satisfy 
the relation $\mu_{n}=\mu_{p}+\mu_{e}-\mu_{\nu_{e}}$
\noindent These situations are characterized by a trapped lepton fraction 
$Y_{L}=Y_{e}+Y_{\nu_{e}}$ 
\noindent where 
$Y_{e}=(n_{e}-n_{e_{+}})/n_{B}$ is the net electron fraction and 
$Y_{\nu_{e}}=(n_{\nu_{e}}-n_{\bar \nu_{e}})/n_{B}$ 
is the net neutrino fraction. 
\par The evolution of a protoneutron star 
begins from a neutrino-trapped situation with $Y_{L}\sim 0.4 $ to one in 
which the net neutrino fraction vanishes and chemical equilibrium without 
neutrinos is established. In this case the chemical equilibrium is modified 
by setting $\mu_{\nu_{e}}=0 $. In all cases, the conditions of charge 
neutrality requires
\begin{equation}
n_{p}=\bar n_{e}
\end{equation}  
\noindent In the nucleon sector, the constants $g_{\sigma N}$, $g_{\omega N} $,
$g_{\rho N} $, b and c are determined by nuclear matter equilibrium 
density $n_{0}=0.16 fm^{-3} $ and the binding energy per nucleon 
$(\sim 16 MeV) $,
the symmetry energy $(\sim 30-35 MeV) $, the compression modulous
$(200 MeV \leq k_{0}\leq 300 MeV)$ and the nucleon Dirac effective mass 
$M^{*}=(0.6-0.7)\times 939 MeV $ at $n_{0}$. Numerical values of the coupling 
constants so chosen are : 
$$\frac{g_{\sigma N}}{m_{\sigma}}=3.434 fm^{-1} :  \frac{g_{\omega N}}
{m_{\omega}}=2.694 fm^{-1} ;  \frac{g_{\rho N}}{m_{\rho}}=2.1  fm^{-1} $$
$$ b=0.00295 , c=-0.00107. $$

\section{Weak Rates and Neutrino Emissivity}

 The dominant mode of energy loss in neutron stars is through neutrino
 emission.The important neutrino emission processes leading to neutron star 
cooling are the so called URCA processes
\begin{equation}
n\longrightarrow p+e^{-}+\bar{\nu}_e \hskip 2 cm p+e^{-}+\longrightarrow 
n+ {\nu}_e 
\end{equation}
\noindent At low temperatures for degenerate nuclear matter, the direct URCA
process can take place only near the fermi energies of participating particles
 and
simultaneous conservation of energy and momentum require the inequality
$$p_{F}(e)+p_{F}(p)\geq p_{F}(n)$$ 
to be satisfied in the absence of the 
magnetic field. This leads
to the well known threshold {\cite{Pethick}} for the proton fraction 
$Y_{p}=\frac{n_{p}}{n_{B}} \geq 11 \% $ thus leading to strong suppression in 
nuclear matter. This condition is satisfied for $n_{B} \geq 1.5 n_{0}$
( where $n_{0}=0.16 fm^{-3}$,is the nuclear saturation density), in a 
relativistic mean field model of interacting n-p-e gas for $B=0$. 
\noindent The standard model of the long term cooling is the modified URCA:
\begin{equation}
(n,p)+n\longrightarrow (n,p)+p+e^{-}+\bar\nu_{e}
~~,~~ (n,p)+p+e^{-}\longrightarrow (n,p)+n+\nu_{e}
\end{equation}
 \noindent which differ from the direct URCA reactions by the presence in the 
initial and final states of a bystander particle whose sole purpose is to 
make possible conservation of momentum for particles close to the Fermi 
surfaces.For weak magnetic field the matrix element for the process remains 
essentially unaffected and the modification 
comes mainly from the phase space factor.  
Treating the nucleons 
non-relativistically and electrons ultra-relativistically, the matrix element squared and summed over spins is given by
\begin{equation}
\sum |M|^2 = 8 G_{F}^2 \cos^2\theta_{c} (4m_{n}^{*} m_{p}^{*}) E_{e} E_{\nu}
		[(1+3g_{A}^2)+(1-g_{A}^2)\cos\theta_{c}]
\end{equation}
\noindent where $g_{A}=1.261$ is the axial-vector coupling constant.
\noindent The emissivity expression is given by
\begin{equation}
\dot{\cal E}_{\nu} =\Bigg[\prod_{i} \int \frac{1}{(2\pi)^3 2E_{i}} \,d^3p_{i}\Bigg] E_{\nu}			\sum\nolimits|M|^2 (2\pi)^4 \delta^{4}(P_{f}-P_{i}) S  
\end{equation}
\noindent where the phase space integrals are to be evaluate over all particle states. The statistical distribution function $ S=f_{n}(1-f_{p})(1-f_{e})$,where $f_{i}'s$ are the Fermi-Dirac distributions.
\noindent We can now evaluate the emissivity in the limit of extreme degeneracy, a situation appropriate in neutron star cores by
 using the standard techniques to perform the phase space 
integrals {\cite{Agoyal}}
\begin{equation}
\dot{\cal E}=\frac{457\pi}{40320} G_{F}^{2}\cos^2\theta_{c}
(1+3g_{A}^{2}) m_{n}^{*}m_{p}^{*} eB T^6 \sum_{\nu=0}^{\nu_{max}}
(2-\delta_{\nu,0})\frac{1} 
{\sqrt{\mu_{e}^2-m_{e}^2-2\nu eB}}
\end{equation}
where
$\nu_{\max}=Int\Bigg(\frac{\mu_{e}^2-m_{e}^2}{2eB}\Bigg)$. 
In the limit of vanishing magnetic field,the sum can be replaced by an integral
and we recover usual expression i.e. the case for B=0.
\begin{equation}
\dot{\cal E}_{\nu}(B=0)=\frac{457\pi}{20160}G_{F}^2 \cos^2\theta_{c}
(1+3g_{A}^2) m_{n}^{*}m_{p}^{*} \mu_{e} T^6 
\end{equation}
\noindent Modified URCA process are considered to be the 
dominant process for neutron 
star cooling. Similarly the energy loss expression with 
appropriate electron phase space, for modified URCA process 
$\dot{\cal E}_{URCA}$ is calculated to be
\begin{eqnarray}
\dot{\cal E}_{URCA}=
&& \frac{11513}{60480}\frac{G_{F}^2\cos^2\theta_{c}}{2\pi}
	g_{A}^2 m_{n}^{*3}m_{p}^{*}\bigg(\frac{f}{m_{\pi}}\bigg)^4
		\alpha_{URCA} T^8	\nonumber  \\ 
&& \sum_{\nu=0}^{\nu_{max}}(2-\delta_{\nu,0})
	\Bigg(\frac{1}{\sqrt{\mu_{e}^{2}-m_{e}^{2}-2\nu eB}}\Bigg)
\end{eqnarray}
\noindent where $f$ is the $\pi-N$ coupling constant $(f^2 \simeq 1)$
and $\alpha_{URCA}$ has  
been estimated to be $=1.54$.
\noindent The above equation in the $B\rightarrow 0$ limit goes over to the standard result{\cite{friman}}
\begin{equation}
\dot{\cal E}_{URCA}(B=0)=\frac{11513}{30240}\frac{G_{F}^2\cos^2\theta_{c}}
	{2\pi}g_{A}^2 m_{n}^{*3}m_{p}
	\bigg(\frac{f}{m\pi}\bigg)^4 \alpha_{URCA}\,\,    \mu_{e}    T^8
\end{equation}
 In the case of super strong magnetic fields such that
$B>B_{c}^{e}$ ($B_c^e=4.41\times 10^{13}$ Gauss)
all electrons occupy the Landau ground state at T=0 which corresponds to
$\nu=0$ state with electron spins pointing in the direction opposite to the magnetic
field. Charge neutrality now forces the degenerate
non-relativistic protons also to occupy the lowest Landau level with proton 
spins pointing in the direction of the field. In this situation we can no longer consider the matrix elements to be
unchanged and they should be evaluated using the exact solutions of Dirac 
equation. Further because nucleons have anomalous magnetic moment, matrix 
elements need to be evaluated for specific spin states separately.
The electron in $\nu=0$ state has energy 
$E_{e}=\sqrt{m_{e}^2+p_{ez}^2}$ and the positive energy spinor 
in $\nu=0$ state is given by
\begin{equation}
 U_{e,-1} = \frac{1}{\sqrt{E_{e}+m_{e}}} \left( 
       		\begin{array}{c}	
		0 \\
		E_{e}+m_{e} \\
		0 \\
		-p_{ez} \\
	     	\end{array}
       		\right)
\end{equation}
Protons are treated non-relativistically
and the energy in $\nu=0$ state is
$E_{p}\simeq \tilde{m_{p}}+\frac{P_{z}^2}{2\tilde{m_{p}}}+U_0^{p}$
\noindent and the non-relativistic spin up operator $U_{p,+1}$ given by  
\begin{equation}
 U_{p,+1}= \frac{1}{\sqrt{2\tilde{m_{p}}}} \left(
					\begin{array}{c}
					\chi_{+} \\
					0 \\
					\end{array}
					\right)
\end{equation}
\noindent where $\tilde m_{p}=m_{p}^*-\kappa_{p}B$. For neutrons we have
\begin{eqnarray}
\Psi_{n,s}(r) &=&\frac{1}{\sqrt{L_{x}L_{y}L_{z}}}e^{-ip_{n,s} \cdot
  r}U_{n,s}(E_{n,s})   \\
 U_{n,s} &=& \frac{1}{\sqrt{2m_{n}^{*}}} \left(
				\begin{array}{c}
					\chi_{s} \\
					0 \\
				\end{array}
				\right)
\end{eqnarray}
and
$E_{n,s} \simeq m_{n}^{*}+\overrightarrow{p_{n}}^2-\kappa_{n} B s+U_0^{n}$
\noindent in the non-relativistic limit.The neutrino wave function is given by
\begin{equation}
\Psi_{\nu}(r)=\frac{1}{\sqrt{L_{x}L_{y}L_{z}}}e^{-ip_{\nu\cdot r}}U_{\nu,s}(E_{\nu})
\end{equation}
Here $U_{\nu,s}$ is the usual free particle spinor, $\chi_{s}$ is the spin 
spinor and the wave function has been normalised in a volume 
$V=L_{x}L_{y}L_{z}$. Using the explicit spinors given above we can now calculate the matrix element
 squared and summed over neutrino states to get 
\begin{equation}
\sum\nolimits |M_{+}^{\dagger}M_{+}|=|K_{+}|^{2}4(4
		m_{n}^{*}\tilde{m_{p}})(1+g_{A})^2(E_{e}+p_{ez})
		(E_{\nu+p_{\nu z}})
\end{equation}
and
\begin{equation}
\sum\nolimits |M_{-}^{\dagger}M_{-}|=|K_{-}|^{2} 16 (4 m_{n}^{*}\tilde{m_{p}})
g_{A}^2(E_{e}+p_{ez})(E_{\nu}-p_{\nu z})
\end{equation}
\noindent The neutrino emissivity is calculated by using the standard techniques for degenerate matter and we get \cite{} 
\begin{eqnarray}
\dot{\cal E}=
&&	\frac{457\pi}{40320} G_{F}^2\cos^2\theta_{c} eB 
	\frac{\tilde{m_{p}}m_{n}^{*}}{p_{F}(e)} T^6 
	\Bigg[\frac{(1+g_{A})^2}{2}
	\theta\bigg(p_{F}^{2}(n,+)\bigg) 
	exp\bigg(\frac{-p_{F}^{2}(n,+)}{2eB}\bigg)\nonumber \\
&&	+\theta\bigg(p_{F}^{2}(n,+)-4p_{F}^{2}(e)\bigg) 
	exp\bigg(\frac{-(p_{F}^{2}(n,+)-4p_{F}^{2}(e))}{2eB}\bigg)\nonumber\\
&&	+ 2g_{A}^{2}{p_{F}(n,+)\longrightarrow p_{F}(n,-)}\Bigg]
\end{eqnarray}
\noindent where $p_{f}(n,+)$ and $p_{F}(n,-)$ are the neutron Fermi
momenta for spins along and opposite to the magnetic field
direction respectively and are given by  
\begin{equation}
\frac{p_{F}^{2}(n,\pm)}{2m_{n}^{*}}=\mu_{n}-m_{n}^{*}-U_{0}^{n} \pm \frac{\kappa_{n}B}{2m_{n}^{*}}
\end{equation}
\indent We thus see as advertised that in the presence of quantising magnetic
 field the inequality $p_{F}(e)+p_{F}(p)\geq p_{f}(n)$ is no longer required
 to be satisfied for the process to proceed, regardless of the value of proton
 fraction and we get non zero energy loss rate.

\section{Bulk Viscosity of Magnetised Neutron Star Matter}
The source 
of bulk viscosity of neutron star matter is the deviation from $\beta$
equilibrium, and the ensuing nonequilibrium reactions, implied by the 
compression and rarefaction of the matter in the pulsating neutron star.
These important reactions are the URCA and the modified URCA processes.
Since the source of bulk viscosity is the deviation from $\beta$
equilibrium, these reactions are driven by non zero value of 
$\Delta\mu=\mu_{n}-\mu_{p}-\mu_{e}$.
 We calculate the bulk viscosity of neutron star matter in the
 presence of magnetic field for direct URCA processes in the linear
 regime i.e. $\Delta\mu << kT$ .
The bulk viscosity $\zeta$ is defined by{\cite{wang}} 
\begin{equation}
\zeta=2\bigg(\frac{dW}{dt}\bigg)_{av}\frac{1}{v_{0}}
\bigg(\frac{v_{0}}{\Delta v}\bigg)^{2}
	\frac{1}{\omega^{2}} 
\end{equation}			
Here $v_{0}$ is the specific volume of the star in equilibrium 
configuration, ${\mit\Delta v}$ is the amplitude of the periodic 
perturbation with period $\tau=\frac{2\pi}{\omega}$ and
$v(t)=v_{0}+{\mit\Delta v}\sin\bigg(\frac{2\pi t}{\tau}\bigg)$
The quantity $(\frac{dW}{dt})_{av}$ is the mean dissipation rate of 
energy per unit mass and is given by the equation
\begin{equation}
\bigg(\frac{dW}{dt}\bigg)_{av}=-\frac{1}{\tau}\int P(t)\frac{dv}{dt}dt
\end{equation}
The pressure $P(t)$ can be expressed near its equilibrium 
value $P_{0}$, as
\begin{eqnarray}
P(t)=P_{0}+\bigg(\frac{\partial P}{\partial v}\bigg)_{0}\Delta v+
	\bigg(\frac{\partial P}{\partial n_{p}}\bigg)_{0}\Delta n_{p} +
	\bigg(\frac{\partial P}{\partial n_{e}}\bigg)_{0}\Delta n_{e}+	
	\bigg(\frac{\partial P}{\partial n_{n}}\bigg)_{0}\Delta n_{n}
\end{eqnarray}
 The change in the number of neutrons, protons and electrons per unit mass 
over a time interval $(0,t)$ due to  URCA reactions (23) is given by
\begin{equation}
-\Delta n_{n}=\Delta n_{p}=\Delta n_{e}=\int_{0}^{t}\frac{dn_{p}}{dt}dt
\end{equation}
The net rate of production of protons, $\frac{dn_{p}}{dt}$ is given by the 
difference between the rates ${\mit\Gamma_{1}}$ 
and ${\mit\Gamma_{2}}$
of the URCA reactions. At equilibrium the two rates
are obviously equal and the chemical potentials satisfy the equality
$\Delta\mu=\mu_{n}-\mu_{p}-\mu_{e}=0$
A small volume perturbation brings about a small change in the chemical 
potentials and the above inequality is no longer satisfied, now  
${\mit\Delta\mu}$ is not  zero and consequently  the reaction rates  
are no longer equal. The net rate of production of protons will thus 
depend upon the value of ${\mit\Delta\mu}$.
In the linear approximation, $\frac{\Delta\mu}{kT}\ll 1$, the net rate 
 can be written as
\begin{equation}
\frac{dn_{p}}{dt}={\mit\Gamma_{1}}-{\mit\Gamma_{2}}=-\lambda{\mit\Delta\mu}
\end{equation}
 Using the thermodynamic relation
$\frac{\partial P}{\partial n_{i}}=-\frac{\partial \mu_{i}}{\partial v}$
 and employing the above relation we obtain
\begin{equation}
\delta P 
= P(t)-P_{0} = -\frac{\partial(\Delta\mu)}{\partial v}
		\int_{0}^{t}\lambda\Delta\mu(t)dt	
\end{equation}
The change in the chemical potential $\Delta\mu(t)$ arises due to 
a change in the specific volume $\Delta v$ and changes in the 
concentrations of various species, viz, neutrons, protons and electrons. Thus
\begin{eqnarray}
 \Delta\mu(t) &=&  \Delta\mu(0)+\bigg(\frac{\partial\Delta\mu}{\partial v}\bigg)_{0}\Delta v
	+\bigg(\frac{\partial\Delta\mu}{\partial n_{n}}\bigg)_{0}
	\Delta n_{n}	
		+\bigg(\frac{\partial\Delta\mu}{\partial n_{p}}\bigg)_{0} 
         \nonumber \\
&&	\Delta n_{p}
		+\bigg(\frac{\partial\Delta\mu}{\partial n_{e}}\bigg)_{0}\Delta n_{e}
\end{eqnarray}
and  we arrive at the following equations for 
$\Delta\mu$:
\begin{equation}
\frac{d\Delta\mu}{dt}=\omega A \frac{{\mit\Delta v}}{v_{0}}\cos(\omega t)-
			C\lambda\Delta\mu
\end{equation}
where
\begin{eqnarray}
A &=& v_{0}\bigg(\frac{\partial\Delta\mu}{\partial v}\bigg)_{0}
              \nonumber \\
C &=& v_{0}\bigg(\frac{\partial\Delta\mu}{\partial n_{p}}+
	\frac{\partial\Delta\mu}{\partial n_{e}}-
	\frac{\partial\Delta\mu}{\partial n_{n}}\bigg)_{0}
\end{eqnarray}
Since for small perturbations, $\lambda$, A and C are constants, equation 
(46)
can be solved analytically to give
\begin{equation}
\Delta\mu=\frac{\omega A}{\omega^{2}+C^{2}\lambda^{2}}
	\frac{{\mit\Delta v}}{v_{0}}
	\bigg[-C\lambda e^{-C\lambda t}+\omega\sin(\omega t)
	+C\lambda\cos(\omega t)\bigg]
\end{equation}
and we obtain
the following expressions for $\zeta$
\begin{eqnarray}
\zeta=\frac{A^{2}\lambda}{\omega^{2}+C^{2}\lambda^{2}}
	\Bigg[1-\frac{\omega C\lambda}{\pi}
	\frac{1-e^{-C\lambda\tau}}{\omega^{2}+C^{2}\lambda^{2}}\Bigg]
\end{eqnarray}
Given the number densities of the these particle species in terms  of
their respective chemical potentials, one can determine the coefficients 
A and C; given the rates  $\Gamma_{1}$ and $\Gamma_{2}$
for the two URCA processes one can determine $\lambda$ and 
hence $\zeta$ for any given baryon density and temperature.
For weak magnetic field several Landau levels are populated
and the matrix elements remain essentially unchanged and one needs to 
account for the correct phase space factor. For non-relativistic 
degenerate nucleons the decay rate constant $\lambda$ is given by {\cite{jd}}
\begin{eqnarray}
\lambda
&=& \frac{17}{480\pi}G_{F}^{2}\cos_{\theta_{c}}^{2} T^{4}		
	 (1+3g_{A}^{2})eB\, m_{p}^*\,m_{n}^*\,\,
	\theta\bigg(p_{F}(p)+p_{F}(e)-p_{F}(n)\bigg)\nonumber \\
&&  \times \sum_{\nu=0}^{\nu_{max}}[2-\delta_{\nu,0}]
	\frac{1}{\sqrt{\mu_{e}^{2}-m_{e}^{2}-2\nu eB}}	
\end{eqnarray}
For strong magnetic field, the electrons are  forced 
into the lowest
Landau level. 
Using the exact wave functions for protons and electrons 
in the lowest Landau level and carrying out the energy integrals for 
degenerate matter, the decay constant $\lambda$
is given by \cite{jd}
\begin{eqnarray}
\lambda
&=& \frac{17}{960\pi}G_{F}^{2}\cos_{\theta_{c}}^{2} eB
	\frac{m_{n}^* m_{p}^*}{p_{F}(p)} T^{4}
	[4g_{A}^{2}+(g_{v}+g_{A})^{2}]
   \Bigg[exp\bigg(-\frac{[p_{F}^{2}(n)-4p_{F}^{2}(e)]}{2eB}
	\bigg)          \nonumber \\
&& \times \theta\bigg(p_{F}^{2}(n)-4p_{F}^{2}(e)\bigg)\nonumber 
	+ exp\bigg(-\frac{p_{F}^{2}(n)}{2eB}\bigg)
	\theta\bigg(p_{F}(n)\bigg)\Bigg]
\end{eqnarray}
It is clear from above that in the case of completely
polarised electrons and protons 
the direct URCA decay rate always gets a non-zero contribution from 
the second term in the last square bracket irrespective of whether the 
triangular inequality $p_{F}(e)+p_{F}(p)\geq p_{F}(n)$ is satisfied or not.

\section{Neutrino Opacity in Magnetised Hot and dense Nuclear Matter}
 We
calculate neutrino opacity for magnetised, interacting dense nuclear
matter for the following limiting cases: a) nucleons and electrons,
highly degenerate with or without trapped neutrinos, b) non-degenerate
nucleons, degenerate electrons and no trapped neutrinos and finally,
c) when all particles are non-degenerate.
The important neutrino 
interaction processes which contribute to opacity are the neutrino
absorption process
\begin{equation}
\nu_e + n\rightarrow p + e
\end{equation}
and the scattering processes
\begin{eqnarray}
\nu_e + N\rightarrow \nu_e + N \nonumber   \\
\nu_e + e\rightarrow \nu_e + e
\end{eqnarray}	
both of which 
get contributions from charged
as well as neutral current weak interactions. For the general process
\begin{equation}
\nu (p_1) + A(p_2)\rightarrow B(p_3) + l(p_4)
\end{equation}
The cross-section per unit volume of matter or the inverse mean free
path is given by
\begin{equation}
\frac {\sigma (E_1)} {V} = \lambda ^{-1}(E_1) = \frac {1} {2E_1} \prod
_{i=2,3,4} d\rho_i \,\,W_{fi}\,\, f_2(E_2)[1-f_3(E_3)]\left[1-f_4(E_4)\right]
\end{equation}
where $d\rho _i=\frac {d^3p_i} {(2\pi )^32E_i}$ is the density of
states of particles and 
the transition rate $W_{fi}= (2\pi )^{4} 
\delta ^{4} (P_f-P_i)\left |M\right |^2$. \\
Weak Magnetic Field : 	 For weak magnetic fields, several Landau levels are populated and the
matrix element remain essentially unchanged and one
needs to only account for the correct phase space factor.	 We first consider the neutrino-nucleon 
processes. In the
presence of weak magnetic fields, the matrix element squared and summed
over initial and final spins in the approximation of treating nucleons
non-relativistically and leptons relativistically is given by
\begin{equation}
\sum \left |M\right |^2 =
32G_{F}^{2}\cos_{\theta_{c}}^{2}m_p^{\star}m_n^{\star}\bigg[(C_{V}^{2} + 3C_{A}^{2})
+ (C_V^2 - C_A^2)cos\theta \bigg]E_eE_{\nu }
\end{equation}
where $C_V=$ $g_V=$1, $C_A=$ $g_{A}=$1.23 for the absorption process;
      $C_V=-$1, $C_A=-$1.23 for neutrino scattering on neutrons and
      $C_V=-1+ 4sin^{2}\theta _{w}=$0.08, $C_A=$1.23 for neutrino
      proton scattering.
We now  obtain neutrino cross-sections in the limits
of extreme degeneracy or for non-degenerate matter. \\
Degenerate Matter :   The absorption cross-section for highly degenerate matter can be
calculated by using (79) in (78) by the usual techniques and we get for
small B \cite{goya}
\begin{eqnarray}
&& \frac {\sigma_{A} (E_{\nu },B)} {V} =
  \frac {G^2_F
\cos_{\theta_{c}}^{2}} {8\pi ^3}(g^2_V + 3g^2_A)m^*_pm^*_nT^2 \frac
{\left(\pi^2 + \frac {(E_\nu - \mu_{\nu})^2} {T^2}\right)} {\left(1 +
e^{ \frac {(\mu_{\nu } - E_{\nu} )} {T}}\right)} eB
                  \nonumber \\
&& \Bigg[\theta
\bigg(p_F(p) + p_F(e)   
 p_F(n) - p_F(\nu )\bigg) + \frac {[p_F(p) + p_F(e) - p_F(n) + p_F(\nu
)]} {2E_\nu }  \nonumber \\
&& \Bigg(\theta \bigg(p_F(\nu )
 - \left |p_F(p) + p_F(e) - p_F(n)\right |\bigg)\Bigg)\Bigg]
	\sum _{\nu =0}^{\nu
  _{max}}(2-\delta _{\nu,0})\frac {1} {\sqrt {\mu^2_e - m^2_e -2\nu
  eB}}
\end{eqnarray}
The case of freely streaming, untrapped neutrinos is obtained from
the above equation by putting $\mu_{\nu }=$0 and replacing $\mu_{e}$ by
$(\mu_{e}+E_{\nu})$.
 When the magnetic field is much weaker than the critical field for
protons, only electrons are affected and the neutrino-nucleon
scattering cross-section expression remain unchanged by the magnetic field. 
The numerical values 
however, are modified due to changed chemical composition. 
The cross-sections are given by 
\begin{equation}
 \frac {\sigma _{\nu N}(E_{\nu})} {V} = \frac
{G^2_F\cos_{\theta_{c}}^{2}} {16\pi ^3}(C^2_V + 3C^2_A){m^*_N}^2T^2
\mu_{e} \frac {\left(\pi^2 + \frac {(E_\nu - \mu _\nu)^2}
{T^2}\right)} {\left(1 + exp\frac {(E_\nu - \mu _\nu)} {T}\right)} 
\end{equation} 
If neutrinos are not trapped, we get in the elastic limit
\begin{equation}
\frac {\sigma _{\nu N}(E_\nu)} {V} = \frac
{G^2_F\cos_{\theta_{c}}^{2}} {16\pi ^3}(C^2_V +
3C^2_A){m^*_N}^2T^{2}E_\nu 
\end{equation} 
 The neutrino-electron scattering cross section is
\begin{equation}
 \frac {\sigma _{\nu e}} {V} \simeq \frac
{2G_F^2\cos_{\theta_{c}}^{2}} {3\pi ^3}(C_V^2 + C_A^2)\frac {\mu _e^2
TE_\nu ^2}{1 + e^{-\beta (E_\nu - \mu_\nu )}} 
\end{equation}  
\noindent which in the untrapped regime goes over to
\begin{equation}
\frac {\sigma _{\nu e}} {V} \simeq \frac {2G_F^2\cos_{\theta_{c}}^{2}} {3\pi ^3}(C_V^2 + C_A^2)\mu _e^2 TE_\nu ^2
\end{equation} 
 Non-Degenerate Matter :
We now treat the nucleons to be non-relativistic non-degenerate 
such that $\mu_i/T \ll -1$ and thus the Pauli-blocking factor $1-f_N(E_i)$ 
can be replaced by 1, the electrons are still considered degenerete and 
relativistic.
The various cross-sections are given by
\begin{eqnarray}
m \frac {\sigma_{A} (E_\nu ,B)} {V}&=& \simeq \frac
{G_F^2\cos_{\theta_{c}}^{2}} {2\pi }(C_V^2 + 3C_A^2)n_n (E_\nu+Q)
\frac {1}{1 + e^{-\beta (E_\nu +Q - \mu_e )}}\nonumber \\ 
&& eB\sum _{\nu =0}^{\nu _{max}}(2-\delta _{\nu,0})\frac {1} {\sqrt{(E_{\nu}+Q)^{2} - m_e^2 -2\nu eB}}
\end{eqnarray}
\noindent  where $n_N$ is the nuclear density and $Q=m_n - m_p$.
\begin{equation}
  \frac {\sigma_{\nu N} (E_\nu ,B)} {V} \simeq \frac
{G_F^2\cos_{\theta_{c}}^{2}} {4\pi }(C_V^2 + 3C_A^2)n_NE_\nu ^2  
\end{equation}
If the electrons too are considered non-degenerate we get
\begin{eqnarray}
 \frac {\sigma _{A}} {V} \simeq && \frac {G_F^2\cos_{\theta_{c}}^{2}}
{\pi }(g_V^2 + 3g_A^2)n_NE_\nu ^2  
            \nonumber \\
\frac {\sigma _{\nu N}} {V} \simeq && \frac
{G_F^2\cos_{\theta_{c}}^{2}} {4\pi }(C_V^2 + 3C_A^2)n_NE_\nu ^2  
\end{eqnarray}
\begin{equation}
 \frac {\sigma _{\nu e}} {N} \simeq \frac
{4G_F^2\cos_{\theta_{c}}^{2}} {\pi ^3}T^4E_\nu (C_V + C_A)^2  
\end{equation}
Quantising Magnetic Field : For quantizing magnetic field the square
of the matrix elements  can be  
evaluated in a straight forward way {\cite{goya}} and we get
\begin{eqnarray}
| M |^2_{A} &=& 8G_F^2cos^2\theta
_cm^*_2m^*_3(E_4+p_{4z})\left[(g_V + g_A)^2(E_1 + p_{1z}) + 4g^2_A(E_1
- p_{1z})\right]          \nonumber \\
&& exp\left[ - \frac {1} {2eB}((p_{1x} + p_{2x})^2 + (p_{3x} +
p_{4y})^2)\right] 
           \\
|M|^2_{\nu p } 
&=& 
	16G^2_F\cos_{\theta_{c}}^{2}C^2_A{m^*_2}^2(p_1
	\cdot p_4 + 2p_{1z}p_{4z}) \nonumber \\
&&	exp\left[-\frac {1} {2eB}((p_{4x} + p_{1x})^2 + (p_{4y} - p_{1y})^2)
	\right]
            \\
|M|^2_{\nu e } &=& 16G^2_F\cos_{\theta_{c}}^{2}
\Bigg[(C^2_V + C^2_A)((E_1E_4 + p_{1z}p_{4z})(E_2E_3 + p_{2z}p_{3z}) 
          \nonumber \\
&& - (E_1p_{4z} + E_4p_{1z})  
 (E_2p_{3z} + E_3p_{2z}))   
 + 2C_VC_A((E_1E_4 + p_{1z}p_{4z}) \nonumber \\
&& (E_2p_{3z} + E_3p_{2z}) - E_1p_{4z} +
E_4p_{1z})  
(E_2E_3 + p_{2z}p_{3z}))\Bigg] \nonumber \\
&& exp\left(- \frac {1} {2eB}((p_{4x} - p_{1x})^2 - (p_{4y} -
p_{1y})^2)\right) 
\end{eqnarray}
 The absorption cross-section is now given by
\begin{eqnarray}
\frac {\sigma _{A}(E_1,B)} {V} &=& \frac {1}{2E_1L_x} \int \frac
{d^3p_z} {(2\pi )^32E_2}\int_{-eBL_x/2} ^{eBL_x/2}\int_{-\infty
}^{\infty } \frac {dp_{3y}dp_{3z}} {(2\pi )^22E_3} \nonumber \\
&& \int_{-eBL_x/2} ^{eBL_x/2}\int_{-\infty }^{\infty } \frac
{dp_{4y}dp_{4z}} {(2\pi )^22E_4}(2\pi )^3\delta (P_y)\delta
(P_z)\delta (E)\left |M\right |^2     \nonumber \\
&& f_2(E_2)[1-f_3(E_3)][1-f_4(E_4)]
\end{eqnarray}
The corresponding scattering cross-sections are obtained by interchanging 
the particle 2 with 4.
Performing the integrals with usual techniques we have 
\begin{eqnarray}
\frac {\sigma _{A}(E_1,B)} {V} \simeq && \frac {eB} {2E_1} \frac {1}
{(2\pi )^3} \frac {1} {8}\int dE_2\frac {dp_{3z}} {E_3}\frac {dp_{4z}}
{E_4}\delta (E)\left |M\right |^2     \nonumber \\
&&
f_2(E_2)[1-f_3(E_3)][1-f_4(E_4)]
\end{eqnarray}

\begin{figure}[t]
\vskip -1cm
\begin{center}
\epsfxsize=\textwidth
\epsfbox{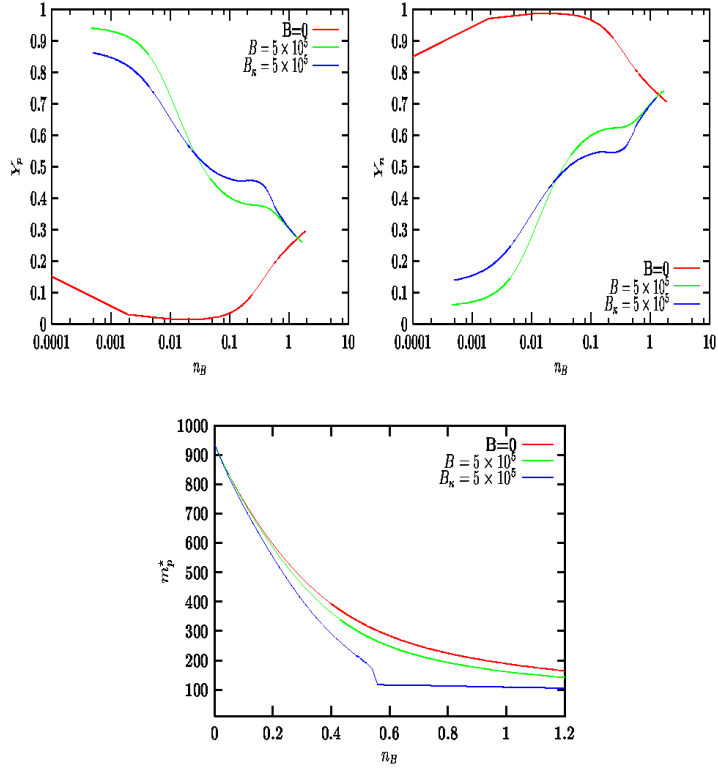}
\caption{Variation of $Y_p$, $Y_n$ and $m_p^*$ with $n_B$ for $B = 0,
5 \times 10^5$ and $B_\kappa = 5 \times 10^5$ where in $B_\kappa$ the
effect of anomalous magnetic moment has been included.}
\end{center}
\end{figure}

To make further progress, we consider the cases of extreme-degeneracy 
and non-degeneracy separately. \\
Degenerate Matter :   For strongly degenerate matter, particles at the top of 
their respective fermi-seas alone contribute  and in this approximation we get \cite{goya}
\begin{eqnarray}
  \frac{\sigma _{A}(E_{\nu},B)}{V}&& 
\simeq   \frac{G^2_Fcos^2\theta _c}{(2\pi )^3}eB
\frac{m^*_nm^*_p}{p_F(\nu )E_\nu }\frac {T^2} {2}\frac {(\pi ^2  
	+ (\frac{E_\nu } {T})^2)} 
	{(1 + e^{-E_\nu /T})} 
	\bigg[(g_V + g_A)^2(E_\nu  + p_{\nu z})  \nonumber \\
&&	+ 4g^2_A(E_\nu  - p_{\nu z})\bigg]
	\Bigg[exp\bigg(-\frac {[p_F(n) + p_F(\nu ) ]^2} {2eB} \bigg) 
	\theta \bigg(p_F(n) + p_F(p ) \bigg)\nonumber \\ 
&&	+exp\bigg(-\frac {[p_F(n) + p_F(\nu )]^2 - 4p^2_F(e)} 
	{2eB}\bigg) \nonumber \\
&& \theta \bigg([p_F(n) + p_F(\nu )]^2 - 4p^2_F(e)\bigg)\Bigg] 
\end{eqnarray}
\noindent The case of freely streaming neutrinos is obtained
by putting $p_{F}(\nu)=0$ everywhere.
\begin{eqnarray}
 \frac {\sigma _{\nu p}(E_\nu ,B)}{V}  
&\simeq& \frac{G^2_Fcos^2\theta _c}{(2\pi )^3}
\frac {{m^*_p}^2}{p^2_F(p)}
g^2_AT\frac {E^2_\nu} {2} \frac{eB}{2}(1 + 2cos^2\theta
)   \nonumber \\ 
 \frac{\sigma _{\nu e}(E_\nu ,B)} {V} &\simeq& \frac{G^2_Fcos^2\theta
_c}{8\pi } T^2eB\mu _\nu \left[(C^2_V + C^2_A)(1 + 
cos^2\theta ) - 4C_VC_Acos\theta \right]  
\end{eqnarray}
 Non-Degenerate Matter: Similarly performing the
integrals in the non degenerate limit, the various cross sections are
\begin{eqnarray}
  \frac {\sigma _{A}(E_{\nu},B)} {V} 
&\simeq&
	\frac {G^2_Fcos^2\theta_c} {4\pi }eB \cos_{\theta}n_{N}
	\frac{1}{e^{-(E_{\nu}+Q-mu_{e})\beta}+1} \nonumber \\
&&  [((g_V + g_A)^2 + 4g^2_A) + ((g_V + g_A)^2 - 4g^2_A)] 
              \\
 \frac {\sigma _{\nu p}(E_\nu ,B)} {V} 
&\simeq& \frac{2G^2_F cos{\theta_{c}}^{2}C^2_A} {2\pi }E^2_\nu n_p 
 \\
 \frac {\sigma _{\nu e}(E_\nu ,B)} {V} 
 &\simeq& \frac {G^2_F} {\pi}n_eTE_\nu\bigg[(C^2_V + C^2_A)(1 +
cos^2\theta )  
- 4C_VC_Acos\theta\bigg ] 
\end{eqnarray}
 We see that in the neutrino free case
($Y_{\nu_{e}}=0)$, the neutrino absorption cross section i.e. the direct 
URCA process which is highly supressed for degenerate matter in the absence 
of magnetic field, proceeds at all densities for the 
quantising magnetic field.

\begin{figure}[h]
\begin{center}
\epsfxsize=6cm
\epsfbox{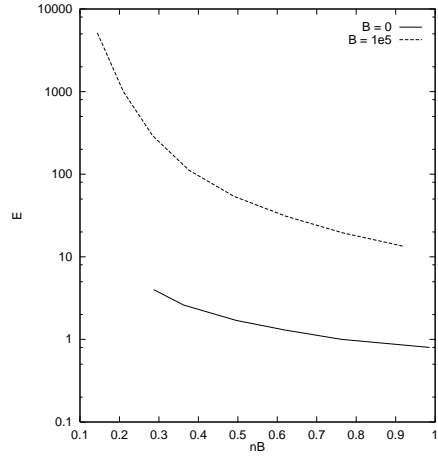}
\caption{Neutrino emmisivity as as function of Baryon density for $B =
0$ and $B = 10^5$, when the electrons and protons are completely
polarized}
\end{center}
\end{figure}

\section{Results and Discussion}

\indent We first calculate the composition of matter for arbitrary
magnetic fields for both the neutrino free and neutrino trapped case
over a wide range of density and temperature. We find that the effect
of magnetic field is to raise the proton fraction and is pronounced at
low densities. The effect of including anomalous magnetic moment also
becomes significant at field strengths $\sim10^5 B_e^c$ which can be
seen from Fig.1 where we have plotted the proton fraction and
effective nucleon mass as a function of density. We then calculate the
effect of magnetic field on neutrino emissivity for the direct URCA
process and find that the effect is not important at weak magnetic
fields. However for the interesting case of quantising magnetic field
capable of totally polarising the electrons and protons, we see from
Fig.2 that in this case the threshold for direct URCA process is
evaded and the emissivity is enhanced by upto two orders of magnitude
and develops anisotropy. 

\begin{figure}
\begin{center}
\epsfxsize=.48\textwidth
\epsfbox{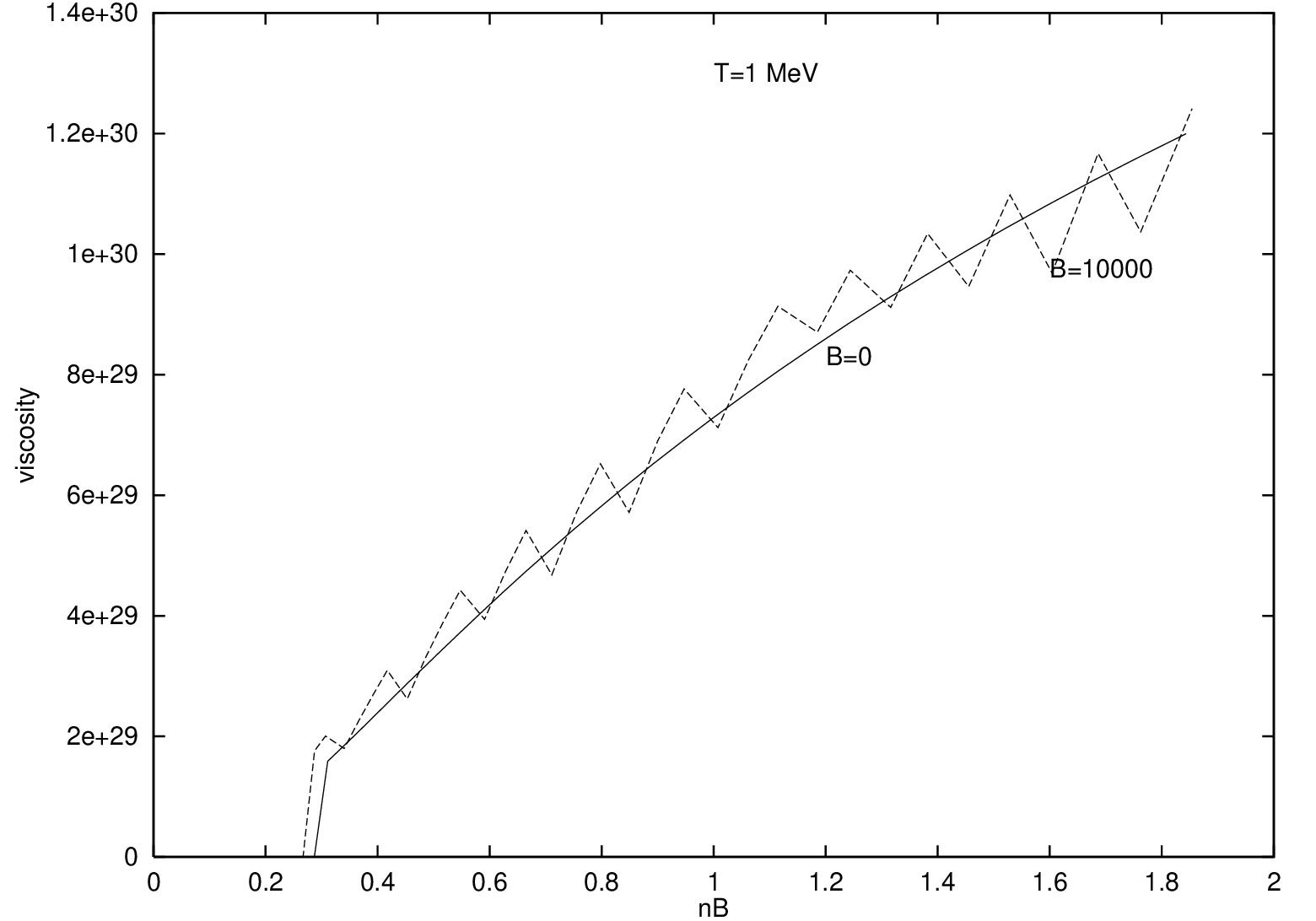}
\epsfxsize=.48\textwidth
\epsfbox{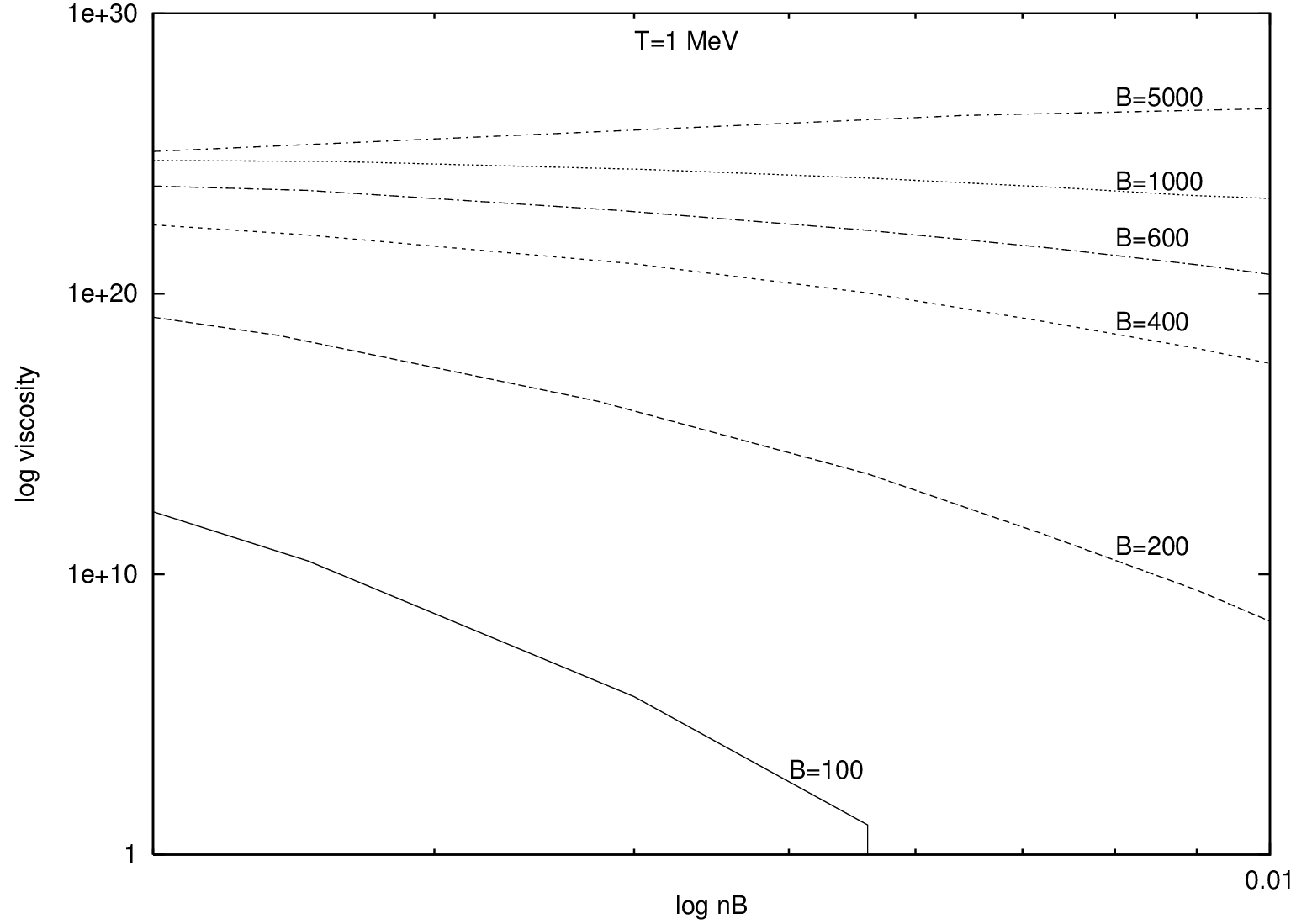}
\caption{Bulk viscosity as a function of Baryon density at $T = 1$ MeV
for different values of magnetic field}
\end{center}
\end{figure}

\begin{figure}[h]
\centering{
\epsfxsize=.48\textwidth
\epsfbox{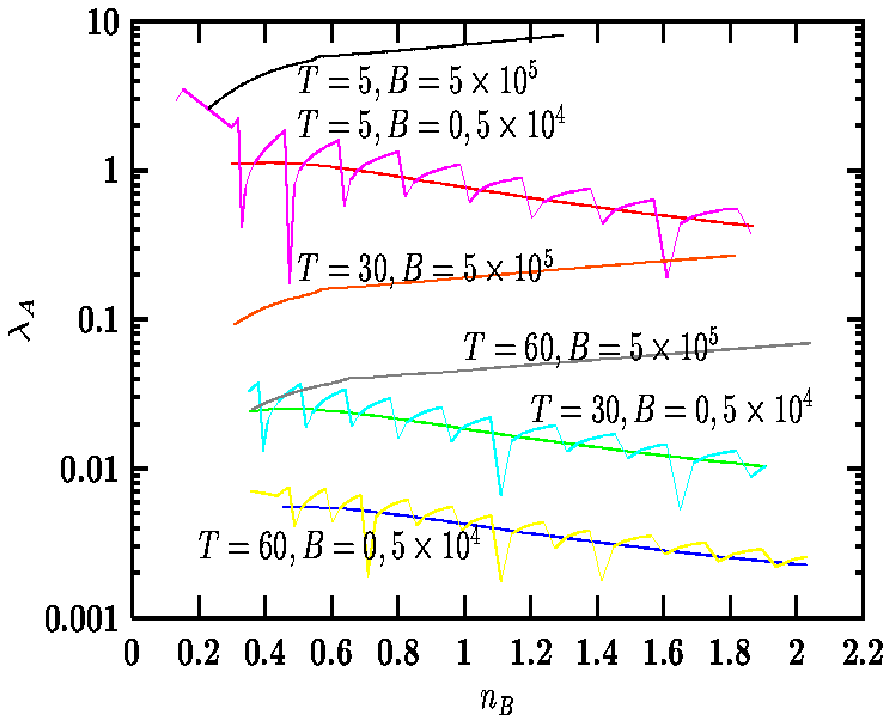}
\epsfxsize=.48\textwidth
\epsfbox{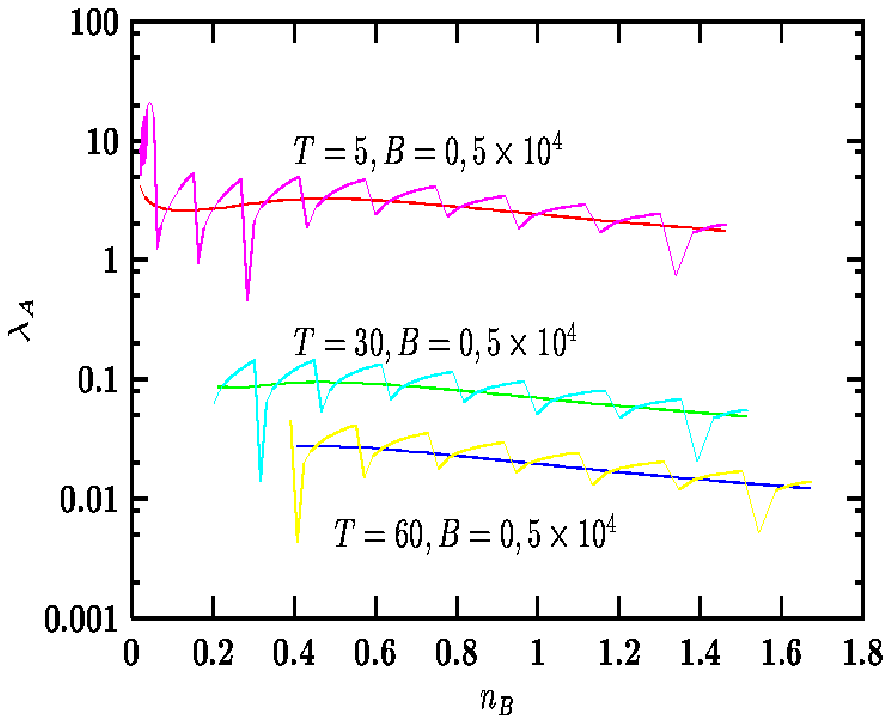}
}
\caption{Variation of absorption mean free path with Baryon
density. The first figure is for untrapped degenerate matter and the
second figure for trapped degenerate matter.}
\end{figure}

$\gamma$-ray burst events if they arise due to pulsations and the
maximum rotation frequency are influnced by the bulk viscosity of
neutron star matter through 1) damping of radial oscillations and 2)
by influencing gravitational radiation reaction instability limiting
maximum rotation rate. In Fig.3 we plot bulk viscosity as a function
of baryon density for different values of magnetic field and find that
the effect of magnetic field is very pronounced at low densities where
only the lowest Landau level contributes and the direct URCA process
is no longer inhibited. We also find that the viscosity decreases very
rapidly with density. 

\begin{figure}
\centering{
\epsfysize=\textwidth
\epsfbox{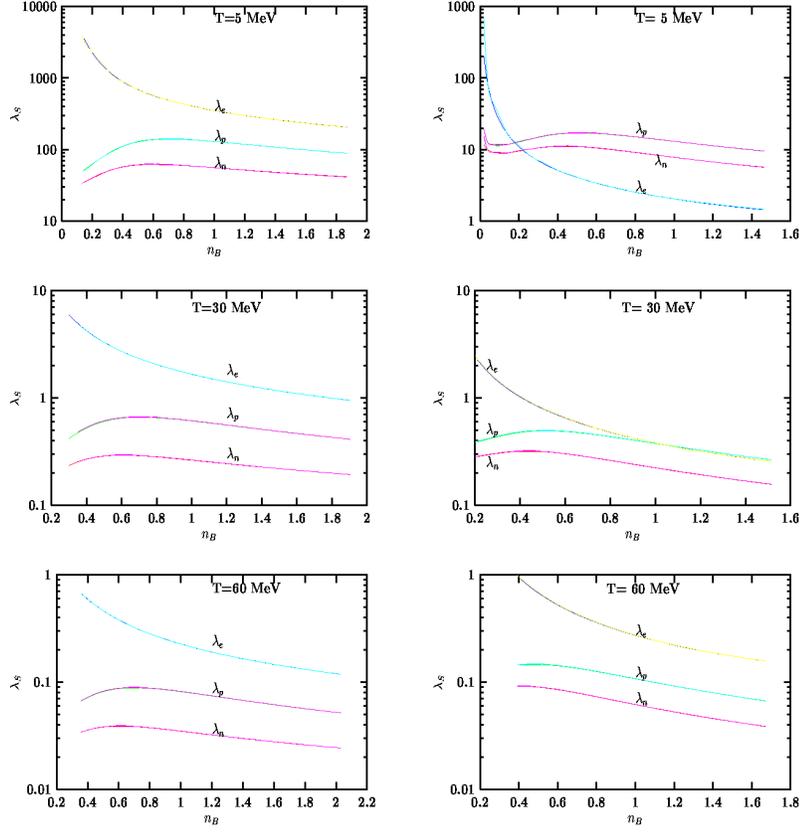}
}
\caption{Variation of neutrino scattering mean free path with baryon
density the left panel is for untrapped matter and the right panel is
for trapped matter}
\end{figure}

Neutrinos emitted during the cooling of protoneutron star have a
momentum $\sim100$ times the momentum of pulsars and therefore, one
percent anisotropy in neutrino emission would give a kick velocity
consistent with observation. This can occur in two different ways
since the neutrino dispersion relations in magnetised medium are
modified. Firstly due to neutrino scattering reactions on polarised
electrons and protons  and secondly {\cite{Vilenk}} due to matter
induced MSW $\nu_e\leftrightarrow\nu_{\tau}$ neutrino oscillations in
the presence of magnetic field when $\nu_{\tau}$ sphere develops
asymmetry and resonant surface gets distorted and the neutrinos
escaping from different depths emerge with different energies. In
Fig.4 and 5 we have plotted the neutrino absorption and scattering
mean free paths with density for different values of temperature and
magnetic field for both the free streaming and trapped regimes. In
Fig.6 we show the variation of absorption mean free path with density
and observe substantial decrease in addition to developing anisotropy
with magnetic field which is particularly pronounced at low
densities. 

To conclude, we find that the effect of quantising magnetic field and
of the inclusion of anomalous magnetic moment of nucleons is to
increase the proton fraction and lower the effective nucleon mass so
as to evade the threshold for the direct URCA processes to
proceed. This results in the enhancement of neutrino emission and bulk
viscosity. This also has the effect of substentially decreasing the
neutrino absorption mean free path in addition to developing
anisotropy.

\begin{figure}
\centering{
\epsfysize=.6\textwidth
\epsfbox{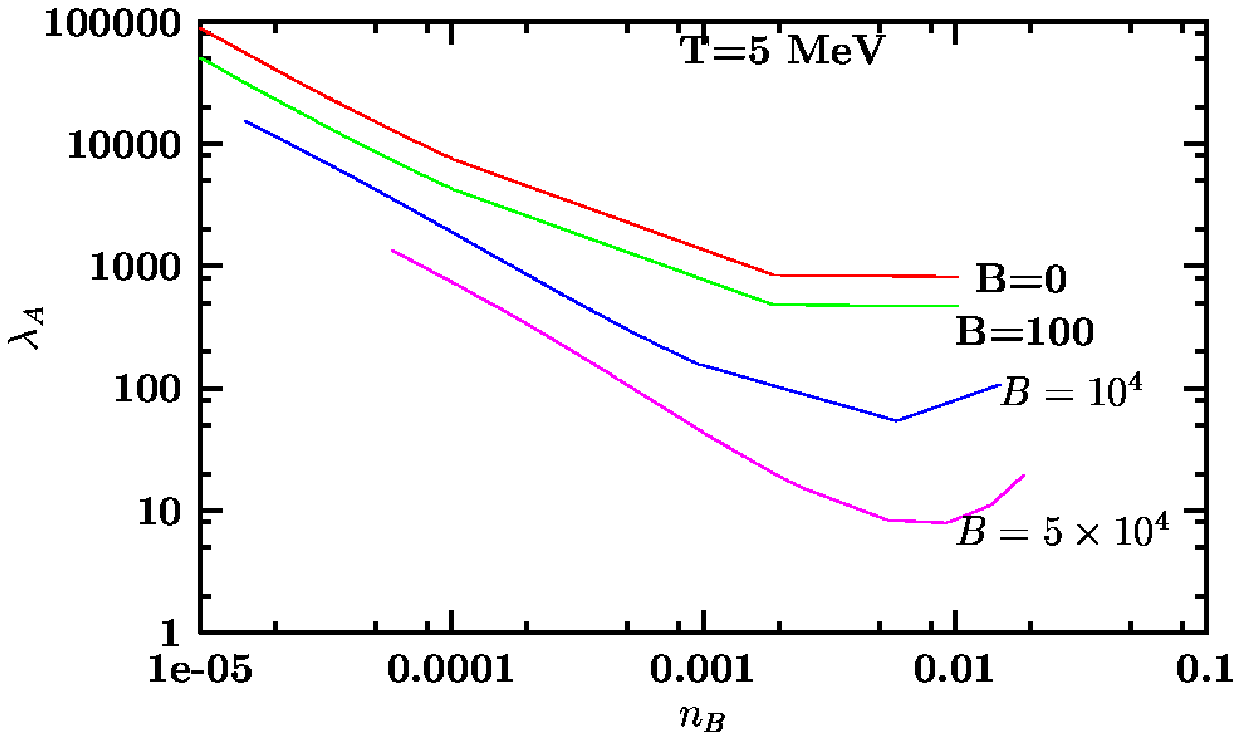}
}
\caption{Variation of neutrino absorption mean free path with Baryon
density for untrapped non-degenerate matter.}
\end{figure}

{\bf{Acknowledgements:}}  I take this opportunity to thank my
coworkers with whom most of this work has been done at various
stages. In particular I thank V.K.Gupta, Deepak Chandra, Kanupriya
Goswami, V.Tuli, J.D.Anand and S.Singh. I thank the organisers of the
workshop for their wonderful hospitality. Partial support from
U.G.C. and D.S.T. (Delhi) is acknowledged.

\end{document}